\documentclass[11pt]{article}
\usepackage{latexsym,amsmath,amssymb,graphics,stmaryrd}
\usepackage{graphics}
\usepackage{subfigure}
\usepackage{fancyhdr}
\usepackage[dvips]{color}
\usepackage{amsfonts}
\usepackage[dvips]{graphicx}
\include{colors}
\def\tr{\,\mbox{Tr}\,}
\newcommand{\intslash}{\backslash {\hskip-2.25ex}\int} 
\begin{document} 
\begin{titlepage}
\renewcommand{\thefootnote}{\fnsymbol{footnote}}
\hfill HU-EP-03/85 ; DFPD 03/TH 48
\vspace*{10mm}
\begin{center}
{\bf \Large Loop Equation in Two-dimensional\\[1mm] Noncommutative Yang-Mills Theory}
\\[20mm]
{\large Harald Dorn\footnote{dorn@physik.hu-berlin.de}
\\Institut f\"ur Physik, Humboldt-Universit\"at zu
Berlin\\Newtonstr. 15, D-12489 Berlin, Germany\\[5mm]
Alessandro Torrielli\footnote{torriell@pd.infn.it}\\Dipartimento di Fisica ``G. Galilei'', INFN Sezione di
Padova\\ Via Marzolo 8, 35131 Padova, Italy}
\vspace*{20mm}

{\bf Abstract}\\[5mm]
\end{center}
The classical analysis of Kazakov and Kostov of the Makeenko-Migdal
loop equation in two-dimensional gauge theory leads to usual partial
differential equations with respect to the areas of windows formed by the
loop. We extend this treatment to the case of $U(N)$ Yang-Mills
defined on the noncommutative plane. We deal with all the subtleties which 
arise in their two-dimensional geometric procedure, using where needed results 
from the perturbative computations of the noncommutative Wilson loop available 
in the literature. The open Wilson line contribution present in the 
non-commutative version of the loop equation drops out in the resulting
usual differential equations. These equations for all $N$ have the same form
as in the commutative case for $N\rightarrow\infty $. However, the
additional supplementary input from factorization properties allowing to solve
the equations in the commutative case is no longer valid.  

\end{titlepage}

\section{Introduction}
\setcounter{footnote}{0}
The loop equation has been the subject of a deep study in the literature,
since the very beginning of its formulation \cite{pol,pol2,migmac,migmac2}.
It is an evolution equation for a class of gauge invariant operators of gauge
theory, the Wilson loop observables, which are considered to be the basic
dynamical objects of the theory: in the confining phase, the knowledge of
these colorless averages determines all the information. The solution of
this loop equation should therefore provide the essential dynamics of the
quantum gauge theory. The right hand side of the equation is given
by an insertion of the Yang-Mills equation of motion operator into the
Wilson loop. The so-called loop operator on the left hand side is constructed
out of suitable second order variational derivatives with respect to the
contour of the Wilson loop. In the formulation
following Polyakov it is given by some projection applied to the
standard second variational derivative, for a more recent version see
\cite{pol3}. In the Makeenko-Migdal approach
it is given by a construction involving the so-called area derivative.

The equation contains singularities, which are connected both
with the usual problem of renormalization of short distance divergences
in multi-point Green functions as well as with the restriction
of D-dimensional distributions to the one-dimensional contour
of the Wilson loop. The precise form of the equation for the renormalized
theory is a delicate issue and the related problems are not completely
solved. There are at least strong arguments that the standard
first and second variational derivation can be interchanged with
the renormalization procedure \cite{dorn}. But due to the short
distance singularities of two insertions for first order variation
\cite{dorn,pol3} the projection recipe \cite{pol3} for the loop operator 
becomes
difficult to handle for practical calculations. On the other side
the area derivative is less well behaved under renormalization.

Gauge theories in two dimensions have a long history serving as a testing area
for problems more complicated in four dimensions. Also in the context of loop
equations the situation becomes better in two dimensions. 
Here a renormalized
version of the Makeenko-Migdal equation has been derived by Kazakov and
Kostov \cite{2}. An important advantage of their equation is
the reduction to an usual partial differential equation with respect
to the areas of the windows formed by a general 2-dimensional loop
with intersections.  They were also able to solve the equation in the large
$N$-limit. In addition for some set of loops solutions have been obtained
for generic $N$, \cite{kaz}.
Remarkably, from their solution one can derive an expression for Wilson loops
winding $n$ times around a single patch \cite{Rossi}, which coincides with the
one obtained via other methods, such as lattice \cite{kkl}, or geometric means
\cite{boul,daul}, or light-cone gauge perturbation theory with 't Hooft's
prescription. In connection with this solution, it was possible to understand
instanton contributions to determine the area-law, and many features of the
light-cone perturbative expansion \cite{Bassetto}.\\

More recently, noncommutative gauge theories have received attention carrying
a new idea of physics at short length, essentially based on their tight
relation to string theory. In this context, new issues arise for gauge
theories, especially the merging of space-time transformation and gauge
symmetries in a larger group, and new classes of instanton solutions. An
important discovery was that in these theories, new gauge invariant
observables come together with the Wilson loops, the so-called noncommutative
open Wilson lines \cite{iikk}. We have therefore a larger set of candidates to encode the
full dynamical content of the theory. The noncommutative
generalization of the Makeenko-Migdal equation was found in \cite{1, kita} for a gauge group $U(N)$

\begin{eqnarray}
\label{1}
\partial^{\mu} {{\delta}\over{\delta \sigma^{\mu \nu}(\xi)}} {{\langle W_c [C]\rangle }\over{V}} &=& - {{g^2 N}\over{V^2}} \int_C d\eta_{\nu}~ \delta^{(D)} (\xi - \eta) \langle W_c [C_{\xi \eta}]\rangle  \langle W_c [C_{\eta \xi}]\rangle  \nonumber \\
&-& {{g^2 N}\over{{(2\pi)}^D V \det \theta}} \int_C d\eta_{\nu} {\langle W_o [C_{\xi
    \eta}] W_o [C_{\eta \xi}]\rangle }_{\mbox{\scriptsize conn}}~.
\end{eqnarray} 
The closed Wilson loop is defined as
\begin{eqnarray}
\label{2} 
W_c [C] = \int dx {{1}\over{N}} \tr U[x + C]~,
\end{eqnarray}
\begin{eqnarray}
\label{3}
U[C] = P_{\star} \exp \Big( i \int_C A_{\mu} (\xi(\tau)) d\xi^{\mu}(\tau)\Big)~,
\end{eqnarray}
while the open lines are
\begin{eqnarray}
\label{4}
W_o [C] = \int dx {{1}\over{N}} \tr U[x + C] \star e^{- i k_{\xi} x}~, 
\end{eqnarray}
where $k_{\xi} = {\theta}^{- 1} (\xi(1) - \xi(0))$ in order to ensure gauge invariance \cite{iikk}. V is the volume of space-time, and we assume the matrix $\theta$ to be nondegenerate.

One realizes that it contains a term which involves also the gauge-invariant open Wilson lines. Moreover, it reproduces the singular nature of the traditional loop equation with the appearance of the delta function in the factorized part, so it is natural to ask if going down to two-dimensions, it will be again possible to give it a regular form. In the noncommutative case the comparison still cannot be performed with exact geometric expressions for the loop, but one can compare for instance with the lattice \cite{bhof,bhof2}, and with the large amount of calculations that have been done in noncommutative light-cone perturbation theory \cite{3,4,5}. Concerning this last case, the results are summarized in the following. 

When using light-cone gauge in two dimensions, there are two different prescriptions for the pole of the propagator that one can use: one is 't Hooft's one \cite{hoo}, which in the commutative case leads to a resummation of the perturbative series of the Wilson loop which coincides with the nonperturbative expression, and which exhibits the area law. Another possible prescription\footnote{The two prescription were put in correspondence with two different ways of quantizing the commutative theory, namely on the light-front ('t Hooft) and at equal-time (Wu-Mandelstam-Leibbrandt or WML) \cite{bbg}. The coordinate space propagators are
\begin{eqnarray}
D_{++}^{'t Hooft} = - {{i}\over{2}} \vert x_+ \vert \delta (x_- ) \, \, , \, \, D_{++}^{WML} = - {{1}\over{2\pi}} {{x_+}\over{(x_- - i \epsilon x_+ )}}~.\nonumber
\end{eqnarray}} 
is WML's one \cite{w,m,le}, which is purely perturbative. The two-dimensional Wilson loop treated with the WML's prescription can be shown to reproduce actually only the zero-instanton sector of the complete result \cite{anlu}, and leads to a resummed expression which consists of exponential times polynomials in the area \cite{stau}. Pure area dependence is ensured by the invariance of the theory under area-preserving diffeomorphisms \cite{wit}. 

In the noncommutative case, the loop computed with 't Hooft's prescription turns out to be ill-defined \cite{3}. The loop computed with WML's prescription is instead finite, with a finite limit different from zero at large noncommutativity and a continuous but non-analytic behaviour when the noncommutativity goes to zero.\\ 

Our aim is to repeat the way to the two-dimensional Kazakov-Kostov version of the
Makeenko-Migdal equation in the light of these results. We will find that it
is still possible, following the same line of reasoning, to arrive at a
regular version of the equation, which is actually
a series of relation among loops in a wide class of different contours. We
will use the input of the explicit Feynman diagrams computations and we will
deal with the subtleties of the derivation in the noncommutative
setting. Remarkably, we realize that in the
final version of the equation the contribution from the open lines is no more
present. Moreover, the final form of the equation turns out to be the same as
the commutative one in the large-$N$ limit, but valid for generic $N$, and
generic $\theta$. At a first glance this could sound quite surprising, since
the perturbative computations show that the noncommutative Wilson loop is
likely to be very different in general from the large $N$ commutative one. The
point is that still the kind of supplementary equation one would write in
order to find an explicit solution should be drastically different. The meaning of the final form of the un-supplemented equation can be to put constraints on the perturbative expansion, either on WML's result or constraining eventual renormalization coefficients introduced in order to give a meaning to
the Wilson loop with 't Hooft propagator. On the other side the equation
provides a relation directly for exact loop
quantum averages, as they can be derived by nonperturbative means.
This opens the perspective of a more powerful test of our equation which
challenges the current available nonperturbative techniques \cite{pasz,go,ambmaknishsz1,ambmaknishsz2,ambmaknishsz3,lucadomenico,bhof,bhof2}. 

\section{Sketch of the Kazakov-Kostov analysis}

Since we will closely follow the classical Kazakov-Kostov derivation of the two-dimensional regular loop equation, we report here for convenience its main points.

The first step is to take the expression for the Wilson loop quantum average
and to perform the area-derivative ${{\delta}/{\delta \sigma^{\mu \nu}(\xi)}}$
(for a detailed explanation, see \cite{intro}). This means one takes the
variation of the loop functional when adding a small loop to the contour with
infinitesimal area at the point $\xi$ on the contour. One then keeps the
contributions linear in the small area. Among these contributions, there is a
first term which reproduces as usual the field-strength via the Stokes
theorem, and that is all on the classical level. In the quantum case and in
the special setting of two dimensions one has a further contribution
instead. Namely,

\begin{eqnarray}
\label{kk1}
&&\delta W_c [C] = {{i}\over{2N}} \langle \tr P \Big[ F_{\mu \nu} (\xi ) \exp \Big( i \int_C A_{\mu} (\xi(\tau)) \, d\xi^{\mu} (\tau) \Big) \Big]\rangle  \, \delta \sigma^{\mu \nu} \nonumber \\
&&-{{1}\over{2N}}\langle \tr \Big[ {\Big( \int_{\delta C} A_{\mu} (\xi(\tau)) \, d\xi^{\mu} (\tau) \Big)}^2 P \exp \Big( i \int_C A_{\mu} (\xi(\tau)) \, d\xi^{\mu} (\tau) \Big) \Big]\rangle \nonumber \\
&&+ {\cal{O}} ({\delta \sigma^2} )~.
\end{eqnarray} 
Computing the contribution to the variation coming from the two point function
integrated around the small added loop as in (\ref{kk1}), see fig.1, one finds a linear dependence on $\delta \sigma^{\mu \nu}$, which is peculiar to two dimensions.  

\begin{figure}
\label{fig1}
\centerline{\includegraphics[width=2.45cm]{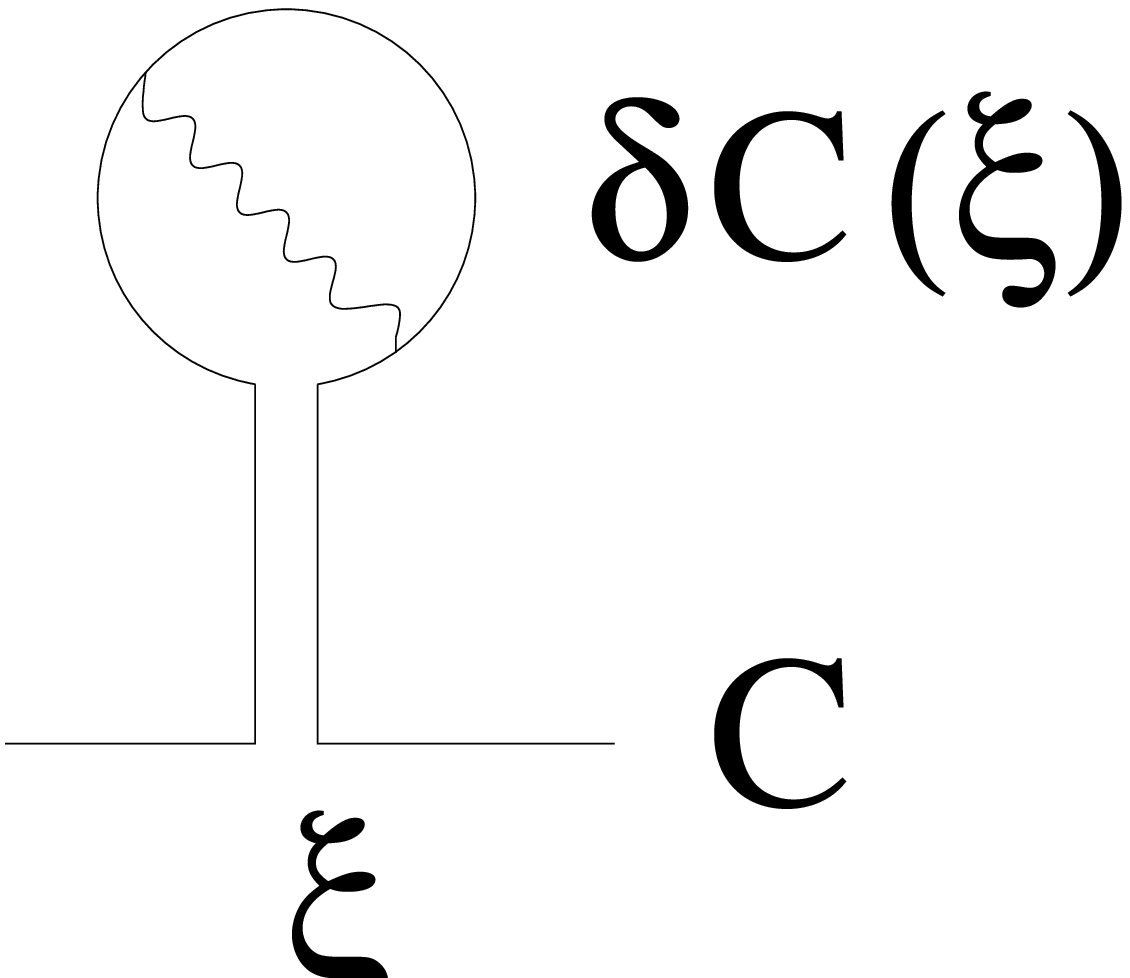}}
\caption{A single propagator inserted in the small loop}
\end{figure}
Then, one takes the point-derivative $\partial^{\mu}$ with respect to the position $\xi$ of the previous insertion. The first term in (\ref{kk1}) turns as usual into the covariant derivative of the field-strength, and it is the way in which one can insert the quantum dynamics: using the quantum equation of motion one turns this term into an integral along the contour of a two-dimensional delta function, times the correlator of the two parts in which the contour is divided by the insertion point $\xi$ and an integration point $\eta$:
\begin{eqnarray}
\label{kk2}
- g^2 N \int_C d\eta_{\nu} \, \delta^{(2)} (\xi(\tau_0 ) - \eta(\tau ))~ \langle W_c [C_{\xi \eta}] W_c [C_{\eta \xi}]\rangle ~. 
\end{eqnarray}
The extra term from (\ref{kk1}) contains now the derivative of the normalized
tensor which defines the small area $\delta \sigma^{\mu \nu}$, which is
basically a sign when changing the orientation; therefore it produces a delta
function contribution. This exactly cancels the contribution coming from the
first term (\ref{kk2}) at the trivial coincidence point $\tau = \tau_0$. What
remains is the contribution in (\ref{kk2}) due to the other value (we assume
unique) of the parameter $\tau$ for which it happens that $\xi(\tau_0 ) =
\eta(\tau )$. This residual term after integration along the contour still
contains a one-dimensional delta function with respect to the orthogonal
direction to the contour. To produce a finite expression one integrates along
a small path of infinitesimal length $2\Delta$ intersecting the contour $C$ at
$\xi$. This results in the equation

\begin{eqnarray}
\label{kk3}
&&\lim_{\vert \Delta \vert \to 0} [\delta / \vert \delta \sigma  (\xi + \Delta
)\vert  - \delta / \vert \delta \sigma  (\xi -\Delta )\vert ] ~\langle W_c [C]\rangle  \nonumber \\
&&= - g^2 N ~\langle W_c [C_1] W_c [C_2]\rangle ~, 
\end{eqnarray}
if $\xi$ is a (simple) self-intersection point and $C_1$ and $C_2$ are the two closed parts in which it divides $C$. If $\xi$ is not a self-intersection point, the r.h.s of (\ref{kk3}) is zero. When taking the large-$N$ limit, the correlator of the two loops in (\ref{kk3}) factorizes in the product of two single quantum averages.

\begin{figure}
\label{fig2}
\centerline{\includegraphics[width=10cm]{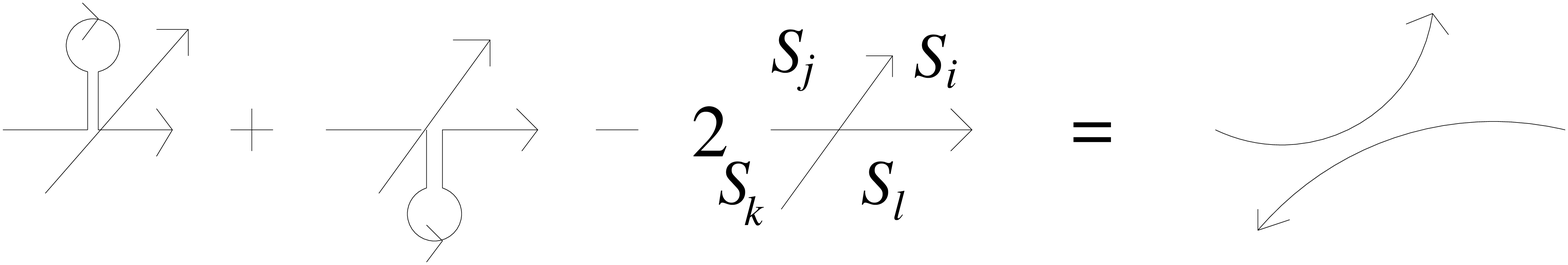}}
\caption{Graphical representation of the final equation}
\end{figure}

Eq.(\ref{kk3}) is now in a regular form: assuming the loop functionals with
many windows to be functions of the areas of these windows alone, due to the
above mentioned symmetry under area-preserving diffeomorphisms, it can be
turned into a system of partial differential equations for these functions, of
the first order in the derivatives with respect to the areas of the windows:

\begin{eqnarray}
\label{kazkos}
\Big( \frac{\partial}{\partial S_k} + \frac{\partial}{\partial S_i} - \frac{\partial}{\partial S_l} - \frac{\partial}{\partial S_j} \Big) ~\langle W_c [C]\rangle  ~=~ - g^2 N \langle W_c [C_1] W_c [C_2]\rangle ~.
\end{eqnarray}
It can be represented graphically as in fig.\ref{fig2}.

Supplementing it with some other assumptions that are justified on the basis of geometrical factorization properties, and on the exponentiation of the simple loop with one window, in the limit of large $N$ (when $\langle W_c [C_1] W_c [C_2]\rangle  = \langle W_c [C_1]\rangle \langle W_c [C_2]\rangle $) one can completely determine the system \cite{2}, and the results reproduce 't Hooft's resummed perturbation theory. These factorization assumptions are not valid for WML's prescription, for which perhaps different supplementary equations could work together with (\ref{kazkos}). 

\section{The Noncommutative Computation}

In this section, we repeat the kind of analysis that Kazakov-Kostov did now in the noncommutative setting. We still assume that acting on the noncommutative Wilson loop with $\partial^{\mu} {{\delta}\over{\delta \sigma^{\mu \nu}}}$ is a well defined operation (in other words, that it is a functional of the Stokes type). In the commutative case this was guaranteed by the pure area dependence due to invariance under area-preserving diffeomorphisms. This survives in the noncommutative setting. Supports to this assumption come from the fact that for example all the explicit computations performed in \cite{3,4,5} show that the Wilson loop averages depend only on the area of the contour. Invariance under area-preserving diffeomorphisms is actually believed to be one of the main ingredient of the structure of the noncommutative gauge group in two dimensions \cite{sh,FedAleSza,ps}.

We will use the light-cone gauge fixing $A_- = 0$ and allow in parallel both the WML's and 't Hooft's prescription for the propagator. In \cite{3} (see also \cite{5}) it was shown that 't Hooft's prescription leads to a series of problems originating from the occurrence of products of distributions which are ill-defined, and resulting in some divergences in the perturbative expansion. It was thought that the origin of this fact, when compared with the finite commutative result, and with the smooth expression one gets using instead WML's prescription as shown in \cite{3,4,5}, could be perhaps attributed to an incompatibility of the punctual interaction that characterises 't Hooft's propagator with the noncommutativity of the space-time. 
All the evidence was then towards avoiding 't Hooft's prescription. Instead, we will take for it another point of view: we take the approach of renormalization of singularities. We therefore assume that the Wilson loop, when treated with 't Hooft's prescription, is however regularized in some way in order to have a finite result, and also renormalized through some subtraction procedure, which leaves it of the Stokes type. 

\subsection{Variation of the Wilson Loop}

In repeating the Kazakov-Kostov analysis we start evaluating the area variation of the functional $W_c [C]$. This means we write

\begin{eqnarray}
\label{5}
{{1}\over{V}} W_c [C\delta C] &=& {{1}\over{V}} \langle  \int dx {{1}\over{N}} \tr P_{\star} \Big[ \exp \Big( i \int_{\delta C} A_{\mu} (x + \xi(\tau)) d\xi^{\mu}(\tau)\Big) \nonumber \\
&&\star \, \exp \Big( i \int_C A_{\mu} (x + \xi(\tau)) d\xi^{\mu}(\tau)\Big) \Big] \rangle ~,
\end{eqnarray}
where $\delta C$ is a small loop of area $\delta \sigma^{\mu \nu}$ we add to
 the original contour $C$ at the point $\xi$ in such a way that drawing
 $\delta C$ does not add new self-intersections \cite{2}. We will consider the
 same class of contours considered in \cite{2} i.e. with an arbitrary number
 of simple self-intersections. According to the previous considerations, we
 still assume that the
corresponding Wilson loops  depend only on the areas of the windows in which the various self-intersections divide the contour. Expanding the contribution from the
$\delta C$-integration in powers of the gauge field as in \cite{2}, we single
out a first term that reproduces the original Wilson loop, a second term
linear in the gauge field on $\delta C$
which is treated in the usual way making use of Stokes' theorem, and then a
tower of contributions containing one or more propagators inside the small
loop $\delta C$ or more than two propagators connecting $\delta C$ with $C$.
Among these, we should keep only the ones at most linear in $\delta \sigma^{\mu \nu}$. In the commutative case this amounts to a single term coming from the two-point function of gauge fields both on the small loop. The corresponding term is now with analogous considerations

\begin{eqnarray}
\label{6}
-~{{1}\over{2V}} \int dx {{1}\over{N}} \tr \Big[ \langle \int_{\delta C}
\int_{\delta C}  A_{\mu} (x + \xi(\tau)) \star A_{\nu} (x +
\xi(\tau'))~d\xi^{\mu}(\tau) d\xi^{\nu}(\tau')\rangle  \nonumber\\
\star \, \langle P_{\star} \exp \Big( i \int_{C} A_{\mu} (x + \xi(\tau)) 
d\xi^{\mu}(\tau)\Big) \rangle \Big]~.~~~~~
\end{eqnarray} 
For this term one can use that quantum averages are indeed invariant under
translations \cite{1}, therefore the two quantum averages in Eq.(\ref{6}) are
independent of $x$. We can bring the first one outside the integral over
$dx$. Moreover, this first quantum average is unaffected by the
noncommutativity, which starts acting when diagrams with crossing propagators
enter the game \cite{3,5}. Using also that the two-point correlator of the
gauge field is diagonal in color space, we see that (\ref{6}) amounts to the
product of ${{1}\over{V}} \langle W_c [C]\rangle $ times the same coefficient
$( - {{1}\over{2}}) \int_{\delta C} \int_{\delta C} D_{++} (\xi(\tau) -
\xi(\tau'))$ as in the commutative case. There, this was the only contribution
linear in the small area $\vert \delta \sigma \vert$ beyond the $F$-insertion
term. Kazakov and Kostov used in particular that Green's function with more
than two gauge fields on the small contour yield
contributions higher order in $\delta \sigma^{\mu \nu}$. 

This property is still true in the noncommutative case for WML \cite{4,5}, but is no more true for 't Hooft. Explicit calculations in \cite{3} have shown that in the contribution from the crossed part of the four-point function to a loop with a single window of area $A$ evaluated with 't Hooft's propagator, one gets a finite part that can be expanded in positive powers of $1 / \theta$, in which each coefficient is proportional to a power of the area larger than one, plus two terms with a positive power of $\theta$, of the form $a \theta + b \theta^2$. These two terms were the origin of the ill-definiteness we mentioned at the beginning, and we assume here them to be suitably regularized and renormalized. A natural assumption is that this can be done without breaking the pure area dependence, which is a requirement that the symmetry under area-preserving diffeomorphism does not produce anomalies\footnote{This procedure could also be interpreted in relation to the renormalization of the field theory of the one dimensional fermion living on the contour, according to the alternative formulation of the Wilson loop vacuum expectation value already known in the commutative case (see e.g. \cite{dorn}), and extended to the noncommutative case for example in \cite{szf}.}. At the end, just by dimensional analysis one expects $b$ to be a function $f_b (A,\lambda)$ of a dimensionless combination of the area $A$ and of some renormalization scale $\lambda$, and $a$ to be given by the area $A$ itself times another function $f_a (A,\lambda)$ with dimensionless combinations. Altogether 

\begin{eqnarray}
\label{definab}
a\theta + b \theta^2 = A f_a (A,\lambda ) \theta + f_b (A,\lambda ) \theta^2.
\end{eqnarray}
These terms can in principle provide a linear dependence on the area.\\

We now want to make a series of important remarks. The assumption that the
symmetry under area-preserving diffeomorphism is not anomalous implies that
one can find at least one procedure of regularization and renormalization such
that it is preserved. We have been able to find a suitable regularization in
coordinate space on a rectangular contour that makes these singular terms to
be actually zero, due to a loop coordinate symmetry. This is in contrast to previous treatments based on a
momentum space procedure \cite{3,4,5}. But as usual different treatments of
such singular distributional integrals can give different results.
We have arguments that this coordinate space procedure can work also at higher
orders, but we are not able to put it into a theorem. However, this strongly
suggests that a regularization exists that does not break at least the pure
area-dependence, and in particular with this regularization $f_a$ and $f_b$
and analog terms from   
higher than four point functions can be zero. 

In our subsequent treatment, since the regularization of this integral is such a delicate issue, we are general and allow for non-vanishing coefficients, also in order to avoid being too tight to the rectangle for these singular terms. The main point we stress here is the following: we will see that whenever non-zero, they however will not change the final form of the equation, due to a geometric constraint we will find below. The form (\ref{final}) and the equations derived from it are stable against this ambiguity.

\subsection{Derivation of the Equation}

Green's functions with two and in principle also more propagators on the small
loop $\delta C$ can potentially provide further contributions linear in the
small area $\vert \delta \sigma \vert$. In order to be general, we will
  write therefore the total extra contribution (i.e. additional to the field strength
  insertion) to the derivative
  ${{1}\over{V}} \partial^{\mu} {{\delta}\over{\delta \sigma^{\mu \nu}}} \langle W_c
  [C]\rangle $ as

\begin{eqnarray}
\label{extra}
{{1}\over{\vert \delta \sigma \vert}} {\Bigg[ {{\langle W_c [\delta C]\rangle }\over{V}} - 1
  \Bigg]}_{\mbox{\scriptsize lin. in $\vert \delta \sigma \vert $
}} \partial^{\mu} n_{\mu \nu} \, {{\langle W_c [C]\rangle }\over{V}}~,      
\end{eqnarray}
where we have introduced a two-dimensional tensor such that ${{\delta}\over{\delta \sigma^{\mu \nu}}} = n_{\mu \nu} {{\delta}\over{\vert \delta \sigma \vert}}$, and we have used again that quantum averages are invariant under translations\footnote{We still assume that more complicated correlators of the small loop with $C$ are higher orders in $\delta \sigma^{\mu \nu}$. Further arguments from the consistency condition (\ref{consist}) in the following indicates they however would not contribute to the final form (\ref{final}) of the equation.}. 

The next step is to use the result of \cite{1} for the term corresponding to the field-strength insertion, namely the right-hand side of Eq.(\ref{1}). We see that this contains a delta-like term similar to the one which commonly appears in the right-hand side of the commutative Makeenko-Migdal equation, but already factorized in the two quantum averages relative to the two sections in which the loop is divided, plus a term which contains the connected correlator of the two Wilson lines corresponding to these sections. We will consider later the piece with the Wilson lines. Restricting to the other term, it has for any generic $N$ the same form as the commutative case at $N \to \infty$. If we write it explicitly in our case it is similar to (\ref{kk2})

\begin{eqnarray}
\label{cancel}
- {{g^2 N}\over{V^2}} \int_C d\eta_{\nu} (\tau ) ~\delta^{(2)} (\xi(\tau_0 ) -
 \eta(\tau)) \langle W_c [C_{\xi \eta}]\rangle  \langle W_c [C_{\eta
 \xi}]\rangle ~. 
\end{eqnarray}   
If $\xi$ is not a self-intersection point, a contribution to (\ref{cancel}) comes only from the trivial coincidence $\tau = \tau_0$. Otherwise there will be other contributions from the other values of the parameter $\tau$ such that $\eta(\tau) = \xi(\tau_0 )$. As in the commutative case, we can conclude that here again (\ref{6}) cancels the trivial coincidence contribution, both for 't Hooft and WML, since the two-point contribution is the same for simple loops. Therefore we write 

\begin{eqnarray}
\label{10}
&{{1}\over{V}} \partial^{\mu} {{\delta}\over{\delta \sigma^{\mu \nu}(\xi )}}& \langle W_c [C]\rangle ~=\nonumber\\ 
&& -~ {{g^2 N}\over{V^2}}~ \intslash d\eta_{\nu}(\tau )~ \delta^{(2)} (\xi
(\tau _0) - \eta (\tau ) \langle W_c [C_{\xi \eta}]\rangle  \langle W_c [C_{\eta \xi}]\rangle  \nonumber \\
&&+ \Bigg[ {{\langle W_c [\delta C]\rangle }\over{V}} \Bigg ]_{\diamondsuit}~{{\langle W_c [C]\rangle }\over{V}}~~\partial^{\mu} n_{\mu \nu}~ + ~\mbox{Wilson lines}~,  
\end{eqnarray} 
where we have defined 

\begin{eqnarray}
\label{exotic}
{\Bigg[ {{\langle W_c [\delta C]\rangle }\over{V}} \Bigg]}_{\diamondsuit} =
{{1}\over{\vert \delta \sigma \vert}} {\Bigg[ {{\langle W_c [\delta C]\rangle }\over{V}}
  \Bigg]}_{\mbox{\scriptsize lin. in $\vert \delta \sigma \vert $,  4 p.t and more}}~.
\end{eqnarray}
The backslash on the integral in  (\ref{10}) indicates the
  omission of a small part of the contour around $\tau = \tau_0$. Note that
$\partial^{\mu} n_{\mu \nu}$ can be rewritten as $\int_{\tau=\tau_0 -
  \epsilon}^{\tau_0 + \epsilon} d\eta_{\nu}(\tau ) \delta^{(2)} (\xi(\tau_0 ) -
\eta(\tau))$. The corresponding term with the two-point function in the small
loop was just responsible for the cancellation of the trivial coincidence
contribution in the first term of the r.h.s. of (\ref{10}).  

We will now require that after renormalization the Wilson loop
in 't Hooft's prescription still fulfills the basic requirement of
approaching one for vanishing enclosed area.
Applying this to the simple loop $\delta C$ that appears in (\ref{10}) implies that the corrections we find from four and higher point functions must approach zero as $\vert \delta \sigma \vert$ goes to zero. Constant or divergent pieces as $\vert \delta \sigma \vert \to 0$, if any, must be set to zero for consistency, and this gives Eq.(\ref{10}) a well defined meaning. In particular, we assume for the functions $f_a$ and $f_b$ previously introduced for the four point function (see eq.(\ref{definab})), the following form:

\begin{eqnarray}
\label{fa}
f_a (A,\lambda) = \sum_{i=0}^{\infty} {(A / \lambda )}^i \, f_a^i \, \, ; \, \, f_b (A,\lambda) = \sum_{i=1}^{\infty} {(A / \lambda )}^i \, f_b^i~.  
\end{eqnarray}
Analogous requirements must be satisfied by higher order Green's functions.\\

At this stage we follow Kazakov and Kostov's strategy of eliminating the surviving delta-like singularities still present in Eq.(\ref{10}). We integrate as in \cite{2} along a small path $\delta C_{\perp}$ intersecting the contour $C$ at the point $\xi(\tau_0)$, and then let the length of this path going to zero. What we get is the following for $\xi$ a self-intersection point: 

\begin{eqnarray}
\label{integr}
&&{{1}\over{V}} \lim_{\vert \Delta \vert \to 0} [\delta / \vert \delta \sigma
(\xi + \Delta )\vert   - \delta / \vert \delta \sigma  (\xi -\Delta )\vert ]~ \langle W_c [C]\rangle  =  \nonumber \\
&&- {{g^2 N}\over{V^2}} \langle W_c [C_1]\rangle  \langle W_c [C_2]\rangle  
~+~  {\Bigg[ {{\langle W_c [\delta C]\rangle }\over{V}} \Bigg]}_{\diamondsuit}~~{{\langle W_c [C]\rangle }\over{V}}  \nonumber \\
&&+ \, \lim_{\vert \Delta \vert \to 0} \int_{\delta C_{\perp }} \mbox {Wilson lines}~, 
\end{eqnarray} 
where $C_1$ and $C_2$ are the two nontrivial sections in which $\xi$ divides
the original contour $C$. We give the operation on the left-hand side the same
meaning as in \cite{2}, and we explicitly translate it into a sum of simple
partial derivatives with respect to the areas $S_m$ of the four windows 
which meet at $\xi $ as in eq.(\ref{kazkos})\footnote{One considers also the external infinite part of the space as a window when $\xi$ is on the external boundary of the loop. In this case, one drops the partial derivative with respect to this area.}.

Now we consider the same for a non self-intersecting point. We get 

\begin{eqnarray}
\label{integrnonself}
{{1}\over{V}} &\lim_{\vert \Delta \vert \to 0}& [\delta / \vert \delta \sigma
(\xi + \Delta )\vert  - \delta / \vert \delta \sigma (\xi -\Delta )\vert ]~ \langle W_c [C]\rangle ~ =  \nonumber \\
&&  {\Bigg[ {{\langle W_c [\delta C]\rangle }\over{V}} \Bigg]}_{\diamondsuit}~~{{\langle W_c [C]\rangle }\over{V}}\nonumber \\&& + ~ \lim_{\vert \Delta \vert \to 0} \int_{\delta C_{\perp }} \mbox{ Wilson 
lines}~. 
\end{eqnarray} 
Since by assumption $\xi $ is a non self-intersecting point, the
left-hand side is zero. This means that we must require for consistency

\begin{eqnarray}
\label{consist}
&& {\Bigg[ {{\langle W_c [\delta C]\rangle }\over{V}} \Bigg]}_{\diamondsuit}{{\langle W_c
    [C]\rangle }\over{V}} + \Bigg[ \, \lim_{\vert \Delta \vert \to 0} \int_{\delta
  C_{\perp }} \mbox{ W. lines} \Bigg]=0~. 
\end{eqnarray} 
In the starting loop equation (\ref{1}) the open Wilson line
contribution, in contrast to the first term on the r.h.s., has no evident
distributional character. Therefore, we expect this term not to contribute in
the limit $\vert \Delta \vert \rightarrow 0$. We comment more on this issue
in the appendix and continue taking the vanishing of the Wilson line
term in (\ref{consist}) for granted. Then also the first term in
(\ref{consist}) has to vanish, i.e. $[\langle W_c [\delta C]\rangle /V]_{\diamondsuit}=0$.
The last statement is independent of the point $\xi $, and using it now in
(\ref{integr}) we get 

\begin{eqnarray}
\label{final}
{{1}\over{V}} \Big ( \frac{\partial}{\partial S_k} + \frac{\partial}{\partial
  S_i} -\frac{\partial}{\partial S_l} -\frac{\partial}{\partial S_j} \Big
)~\langle W_c [C]\rangle ~ = - ~{{g^2 N}\over{V^2}} \langle W_c [C_1]\rangle  \langle W_c [C_2]\rangle ,
\end{eqnarray}
which has the same form as the commutative equation for $N \to \infty$, but it is valid for generic $N$ now\footnote{We stress that the explicit presence of the volumes is in the right form in order to cancel the volume contribution coming from each quantum average due to translational invariance.}.      

Following the line of reasoning of Kazakov and Kostov
 we arrived at a non-singular version of the Makeenko-Migdal equation for the
 noncommutative case in two dimensions. Equation (\ref{final}) is our main
result and is expected to be valid also nonpertubatively.

A first remark is about the solutions of this equation. As already in the
commutative case, as it stands it cannot be solved without implementing some
further input. Kazakov and Kostov used results from explicit calculations with
't Hooft's prescription, and derived a further equation which, combined with
the previous in the large $N$ factorized limit, allowed them to find an
explicit solution within the 't Hooft prescription (corresponding to the exact geometrical
one). We cannot use their supplementary equation because looking at the
computations in \cite{3} we see it is clearly false in the noncommutative
case, and it seems difficult to find at a first sight a valid substitute. 

The second remark concerns the role our equation can play in
constraining otherwise open renormalization constants for the Wilson loop
in 't Hooft prescription. From the discussion of (\ref{consist}) we know
\begin{eqnarray}
\label{consist2}
&&{\Bigg[ {{\langle W_c [\delta C]\rangle }\over{V}} \Bigg]}_{\diamondsuit} \, = \, \, 0~,
\end{eqnarray} 
which can be read as the vanishing of the coefficients $f_a^0$ and $f_b^1$ of
Eq.s (\ref{fa}) and the analog ones from higher point Green functions. We have
therefore that the simple renormalized loop with one window and one winding,
apart from the known two-point function contribution, should have no terms
linear in the area. In addition one can extract from (\ref{final}) information about
the relation among the remaining coefficients of the simple Wilson loop
and the ones for Wilson loops with more than one window.

\section{Loops with Multiple Windings}

As a remarkable example, one can repeat the procedure which allows to derive
from (\ref{final}) an equation for loops with a single window but $n$ winding
around this window. These loops were object of study in the commutative case
as a special important class of contours (see \cite{Rossi,Bassetto}), and in the noncommutative case were studied in \cite{4}. 

\begin{figure}
\label{fig3}
\centerline{\includegraphics[width=2.8cm]{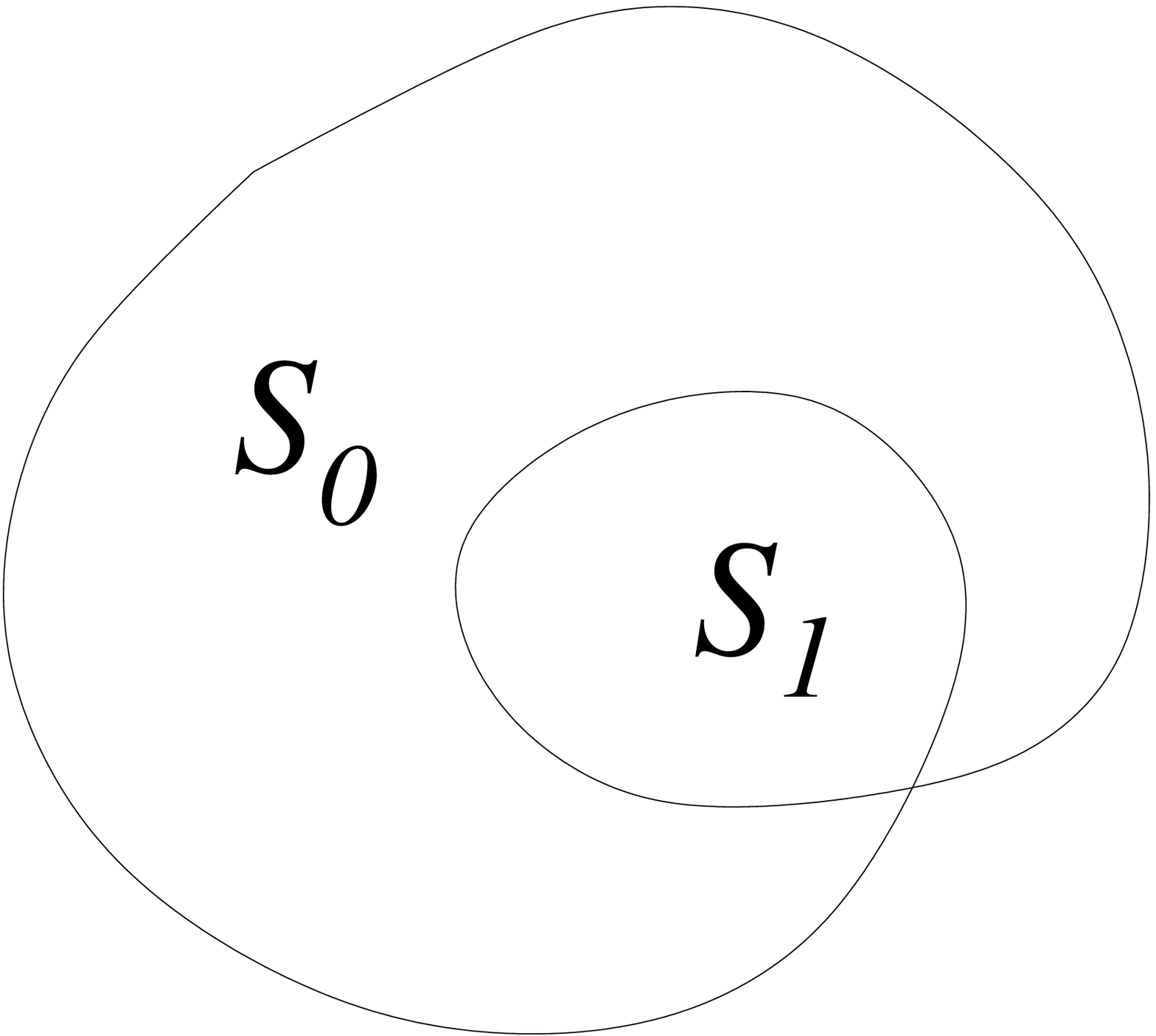}}
\caption{A loop with two windows}
\end{figure}
\noindent
For instance, we can specialise Eq.(\ref{final}) to the situation of a loop
with one simple intersection point which divides it into two windows, one of
area $S_1$ and one of area $S_0$, such that the boundary of $S_1$ is in common
with $S_0$, see fig.3. The idea is that one sends the area $S_0$ to zero and obtains a loop with a single window of area $S_1$ but winding twice around it. Then eq.(\ref{final}) can be written in the following way
(see also footnote 4):

\begin{eqnarray}
\label{final2}
{{1}\over{V}} ( 2 \, \partial / \partial S_0 - \partial / \partial S_1 )~ W_c^{(2)} [S_0 ,S_1 ]~ =~ 
- {{g^2 N}\over{V^2}} W_c^{(1)} [S_1 ] \, W_c^{(1)} [S_0 + S_1 ]~.
\end{eqnarray}
In the commutative case one can supplement this equation with the
  above mentioned further input, which states that the dependence of
$W_c^{(2)}$ on the outer area $S_0$ reduces to a pure exponential, and that
the value of $W_c^{(1)}$ is again a pure exponential of the argument.
Then the equation for $S_0\rightarrow 0$ relates the derivative of $W_c^{(2)}
[ S_1 ]$ to the square of $W_c^{(1)}[ S_1 ]$. Generalizing to higher windings
one was able to find equations closed within the set of loops with one window
but multiple windings.

In the noncommutative case this is no more possible since we cannot use that
particular further input. The dependence on the boundary area of $W_c^{(2)}$ is no more so simple, and certainly the single loop is not an exponential\footnote{This was commented in \cite{3} as something related to the meaning of abelian-like exponentiation as a test for unitarity in the noncommutative case.}. If we still want to find from (\ref{final}) an equation for the loop with two windings on a single window $W_c^{(2)} [S_1 ]$, the form of this equation would be

\begin{eqnarray}
\label{final2wind}
{{1}\over{V}} \, \partial / \partial S_1  W_c^{(2)} [S_1 ]& = &
{{g^2 N}\over{V^2}} W_c^{(1)} [S_1 ] \, W_c^{(1)} [S_1 ]\nonumber\\
& + &{\Big( {{2}\over{V}} \, \partial / \partial S_0 W_c^{(2)} [S_0 , S_1 ]\Big) }_{S_0 = 0}~, 
\end{eqnarray}
which involves informations also from outside the class of winding loops\footnote{In the commutative case, due to the pure exponential dependence mentioned above, the term in round brackets reproduces $W_c^{(2)} [S_1 ]$, and the equation takes a closed structure already reported by Olesen \cite{Olesen}.}.

\section{Conclusions}
We have repeated the analysis of Kazakov and Kostov \cite{2} of the Makeenko-Migdal loop equation in two-dimensional Yang-Mills theory, deriving its noncommutative version on the plane. We started from the results of \cite{1} in general dimensions, and singled out the features peculiar to the case $D=2$. We took results from the light-cone gauge perturbative computations available in the literature, allowing in principle for both 't Hooft's and WML's prescriptions. We realized that extra contributions can arise in the intermediate steps in 't Hooft's case, which however do not contribute to the final form of the equation due to a geometric constraint. We also saw the disappearing of the contribution from the open Wilson lines, found in \cite{1}, after performing the integration along a small path orthogonal to the contour, which is a key component of the procedure. At the end, the equation turns out to have the same form as the commutative one at large $N$, but here is valid for generic $N$, and generic $\theta$. This form is stable both against switching between the 't Hooft and WML prescription and against a series of subtle ambiguities we encounter in its derivation. We interpret the equation both as putting a series of constraints on the perturbative expansion, and as well as the full nonperturbative relation to be fulfilled by the exact expression of the Wilson loop.

We have discussed about the concrete solution of the loop equation, which was possible in the commutative case. Here, the supplementary condition necessary in order to determine the system is much more difficult to find, and we are insofar unable to construct it. We also expect the supplementary input to be different within 't Hooft and WML prescription. In particular, the simple commutative equation describing loops with multiple windings is no more recoverable, and we have shown the new form, that involves information also from outside the class of winding loops. Solving explicitly this equation is still a challenge, especially from the point of view of nonperturbative approaches, and we are going to undertake it in future investigations.   

\section{Acknowledgements}

One of us (A.T.) wishes to thank Antonio Bassetto, Giancarlo De Pol, Federica Vian and Giuseppe Nardelli for very useful and stimulating discussions. He also wishes to thank the ``Deutscher Akademischer Austauschdienst'' (D.A.A.D.) for support under the reference A0339307 (sec.14). The
work of H.D. is supported in part by Deutsche Forschungsgemeinschaft
within the ``Schwerpunktprogramm Stringtheorie'' and by the European
Community Human Potential Programme HPRN-CT-2000-00131 Quantum Spacetime.

\section{Appendix}

We now turn to the problem of disregarding the contribution from the Wilson
 lines in the procedure of integrating along a small path intersecting the
 contour. We already mentioned that this amounts to excluding a certain kind
 of singularities from this contribution, for instance possible delta
 functions which pick up a finite result when $\Delta$ in Eq.(\ref{integr})
 goes to zero. In order to see which kind of behaviour we should expect, we
 have performed some rough computations of the Wilson lines term, using 't
 Hooft's prescription in perturbation theory.
We have analyzed as prototypes some terms from the four point function, splitting as in \cite{3} the propagator in momentum space in the WML's one plus terms involving delta functions. The contribution therefore splits into the smooth WML's result for the lines \cite{BF,BFG}, plus potential singular contributions. Due to the form of the light-cone propagator, these terms can typically produce contributions in Eq.(\ref{integrnonself}) of the form
\begin{eqnarray}
\label{translinv}
\int_C d\eta_{\nu} \lim_{\vert \Delta \vert \to 0}  \int_{\delta C_{\perp }}  \delta ({[k_{\eta \xi}]}_- ) \, \Theta ({[k_{\eta \xi}]}_+ ) \, K(k_{\eta \xi}), 
\end{eqnarray}
where $k_{\eta \xi}=\theta^{-1} (\eta - \xi)$, $\Theta$ is the Heaviside step-function, and $K$ is some function of $k_{\eta \xi}$ one could obtain performing the loop integrations with some regularizing cutoff and subtraction were needed. The direction $\nu$ appearing in (\ref{translinv}) is fixed for the whole integration, and due to the invariance of the two-dimensional theory under area-preserving diffeomorphism, we can always choose it to be along $x_-$. Recalling the definition of $k_{\eta \xi}$ we write therefore
\begin{eqnarray}
\label{translinv2}
\int_C d\eta_{-} \lim_{\vert \Delta \vert \to 0}  \int_{\delta C_{\perp }}  \delta (\eta_- - \xi_- ) \, \Theta (\eta_+ - \xi_+ ) \, K. 
\end{eqnarray}
In this way, the delta function singularity involves only the integration along the contour and not along the small path $\delta C_{\perp}$. The integration procedure and $\Delta \to 0$ thus give a vanishing contribution. Even if we found this argument for the four-point function, the structure of the light-cone integration makes us suppose that at a generic order in perturbation theory the situation will be roughly the same, although in order to have a proof one should rely on a complete analysis.


\begin{thebibliography}{99}

\bibitem{pol}
A.~M.~Polyakov,
Phys.\ Lett.\ B {\bf 82} (1979) 247.


\bibitem{pol2}
A.~M.~Polyakov,
Nucl.\ Phys.\ B {\bf 164} (1980) 171.

\bibitem{migmac}
Y.~M.~Makeenko and A.~A.~Migdal,
Phys.\ Lett.\ B {\bf 88} (1979) 135
[Erratum-ibid.\ B {\bf 89} (1980) 437].

\bibitem{migmac2}
Y.~Makeenko and A.~A.~Migdal,
Nucl.\ Phys.\ B {\bf 188} (1981) 269
[Sov.\ J.\ Nucl.\ Phys.\  {\bf 32} (1980\ YAFIA,32,838-854.1980) 431.1980\ YAFIA,32,838].

\bibitem{pol3}
A.~M.~Polyakov and V.~S.~Rychkov,
Nucl.\ Phys.\ B {\bf 581} (2000) 116
[arXiv:hep-th/0002106].

\bibitem{dorn}
H.~Dorn,
Fortsch.\ Phys.\  {\bf 34} (1986) 11.

\bibitem{2}
V.~A.~Kazakov and I.~K.~Kostov,
Nucl.\ Phys.\ B {\bf 176} (1980) 199.

\bibitem{kaz}
V.~A.~Kazakov,
Nucl.\ Phys.\ B {\bf 179} (1980) 283.

\bibitem{Rossi}
P.~Rossi,
Annals Phys.\  {\bf 132} (1981) 463.

\bibitem{kkl}
V.~A.~Kazakov and I.~K.~Kostov,
Phys.\ Lett.\ B {\bf 105} (1981) 453.

\bibitem{boul}
D.V.~Boulatov, Mod. Phys. Lett. {\bf A9} (1994) 365
[arXiv:hep-th/9310041]

\bibitem{daul}
J-M~Daul and V. A.~Kazakov, Phys. Lett. {\bf B335} (1994) 371
[arXiv:hep-th/9310165]

\bibitem{Bassetto}
A.~Bassetto, L.~Griguolo and F.~Vian,
Nucl.\ Phys.\ B {\bf 559} (1999) 563
[arXiv:hep-th/9906125].

\bibitem{iikk}
N.~Ishibashi, S.~Iso, H.~Kawai, and Y.~Kitazawa,
Nucl.\ Phys.\ B {\bf 573} (2000) 573
[arXiv:hep-th/9910004]

\bibitem{1}
M.~Abou-Zeid and H.~Dorn,
Phys.\ Lett.\ B {\bf 504} (2001) 165
[arXiv:hep-th/0009231].

\bibitem{kita}
A.~Dhar and Y.~Kitazawa,
Nucl.\ Phys.\ B {\bf 613} (2001) 105
[arXiv:hep-th/0104021].

\bibitem{bhof}
W.~Bietenholz, F.~Hofheinz and J.~Nishimura,
JHEP {\bf 0209} (2002) 009
[arXiv:hep-th/0203151].

\bibitem{bhof2}
W.~Bietenholz, F.~Hofheinz and J.~Nishimura,
Nucl.\ Phys.\ Proc.\ Suppl.\  {\bf 119} (2003) 941
[arXiv:hep-lat/0209021].

\bibitem{3}
A.~Bassetto, G.~Nardelli and A.~Torrielli,
Nucl.\ Phys.\ B {\bf 617} (2001) 308
[arXiv:hep-th/0107147].

\bibitem{4}
A.~Bassetto, G.~Nardelli and A.~Torrielli,
Phys.\ Rev.\ D {\bf 66} (2002) 085012
[arXiv:hep-th/0205210].

\bibitem{5}
A.~Torrielli,
PhD Thesis, arXiv:hep-th/0301091.

\bibitem{hoo}
G.~'t Hooft, Nucl.\ Phys.\ {\bf B75} (1974) 461.

\bibitem{bbg}
A.~Bassetto, F.~De Biasio and L.~Griguolo,
Phys.\ Rev.\ Lett.\  {\bf 72} (1994) 3141
[arXiv:hep-th/9402029].

\bibitem{w}
T.T.~Wu, Phys.\ Lett.\ {\bf 71B} (1977) 142.

\bibitem{m}
S.~Mandelstam, Nucl.\
Phys.\ {\bf B213} (1983) 149. 

\bibitem{le}
G.~Leibbrandt, Phys.\ Rev.\ {\bf D29} (1984) 1699.

\bibitem{anlu}
A.~Bassetto and L.~Griguolo,
Phys.\ Lett.\ B {\bf 443} (1998) 325
[arXiv:hep-th/9806037].

\bibitem{stau}
M.~Staudacher and W.~Krauth,
Phys.\ Rev.\ D {\bf 57} (1998) 2456
[arXiv:hep-th/9709101].

\bibitem{wit}
E.~Witten,
Commun.\ Math.\ Phys.\  {\bf 141} (1991) 153.

\bibitem{pasz}
L.~D.~Paniak and R.~J.~Szabo,
JHEP {\bf 0305} (2003) 029
[arXiv:hep-th/0302162].

\bibitem{go}
A.~Gonzalez-Arroyo and M.~Okawa,
Phys.\ Rev.\ D {\bf 27} (1983) 2397.

\bibitem{ambmaknishsz1}
J.~Ambjorn, Y.~M.~Makeenko, J.~Nishimura and R.~J.~Szabo,
JHEP {\bf 9911} (1999) 029
[arXiv:hep-th/9911041].

\bibitem{ambmaknishsz2}
J.~Ambjorn, Y.~M.~Makeenko, J.~Nishimura and R.~J.~Szabo,
Phys.\ Lett.\ B {\bf 480} (2000) 399
[arXiv:hep-th/0002158].

\bibitem{ambmaknishsz3}
J.~Ambjorn, Y.~M.~Makeenko, J.~Nishimura and R.~J.~Szabo,
JHEP {\bf 0005} (2000) 023
[arXiv:hep-th/0004147].

\bibitem{lucadomenico}
L.~Griguolo and D.~Seminara,
arXiv:hep-th/0311041.

\bibitem{intro}
Y.~M.~Makeenko,
Surveys High Energ.\ Phys.\  {\bf 10} (1997) 1.

\bibitem{sh}
M.~M.~Sheikh-Jabbari
JHEP {\bf 9906} (1999) 015
[arXiv:hep-th/9903107]

\bibitem{FedAleSza}
F.~Lizzi, R.~J.~Szabo and A.~Zampini,
JHEP {\bf 0108} (2001) 032
[arXiv:hep-th/0107115].

\bibitem{ps}
L.~D.~Paniak and R.~J.~Szabo,
arXiv:hep-th/0203166

\bibitem{szf}
R.~J.~Szabo,
Phys.\ Rept.\  {\bf 378} (2003) 207
[arXiv:hep-th/0109162].

\bibitem{Olesen}
P.~Olesen,
NBI-HE-81-5
{\it Lectures given at the 18th Winter School of Theoretical Physics, Karpacz, Poland, Feb 18 - Mar 4, 1981}

\bibitem{BF}
A.~Bassetto and F.~Vian,
JHEP {\bf 0210} (2002) 004
[arXiv:hep-th/0207222].

\bibitem{BFG}
A.~Bassetto, G.~De Pol and F.~Vian,
JHEP {\bf 0306} (2003) 051
[arXiv:hep-th/0306017].

\end{thebibliography}
\end{document}